\documentclass[10pt,aps,twocolumn,floats,floatfix,amssymb,prd,superscriptaddress,nofootinbib]{revtex4-2}

\usepackage{graphicx}
\usepackage{amsmath,amssymb}
\usepackage{amsfonts}
\usepackage{xspace} 
\usepackage[usenames]{color}
\usepackage{dcolumn}
\usepackage{bm}
\usepackage{mathrsfs}
\usepackage[colorlinks=true]{hyperref}
\usepackage[all]{hypcap} 
\usepackage[utf8]{inputenc} 
\usepackage{multirow}
\usepackage{etoolbox}
\usepackage{tikz}

\begin{document}

\title{Rotating black holes in semiclassical gravity}

\author{Pedro G. S. Fernandes}
\affiliation{School of Physics and Astronomy, University of Nottingham, University Park, Nottingham, NG7 2RD, United Kingdom}
\begin{abstract}
We present analytic stationary and axially-symmetric black hole solutions to the semiclassical Einstein equations that are sourced by the trace anomaly. We also find that the same spacetime geometry satisfies the field equations of a subset of Horndeski theories featuring a conformally coupled scalar field. We explore various properties of these solutions, and determine the domain of existence of black holes. These black holes display distinctive features, such as non-circularity, a non-spherically symmetric event horizon and violations of the Kerr bound.
\end{abstract}

\maketitle 

%%%%%%%%%%%%%%%%%%%%%%%%%%%%%%%%%
\noindent{\bf{\em Introduction.}}
%\section{Introduction}
%%%%%%%%%%%%%%%%%%%%%%%%%%%%%%%%%
The trace anomaly \cite{Capper:1974ic,Duff:1993wm} is a quantum-level phenomenon related to the breaking of conformal symmetry of a conformally invariant classical theory. It occurs due to one-loop quantum corrections that result in a renormalized stress-energy tensor with expectation value $\langle T_{\mu \nu} \rangle$ and a non-zero trace. This is a general feature of quantum theories in gravitational fields, on the same footing \cite{Balbinot:1999ri,Balbinot:1999vg,Anderson:2007eu,Mottola:2022tcn,Mottola:2006ew,Mottola:2016mpl} as the chiral anomaly in QCD, which led to the successful prediction of the decay rate of the neutral pion into two photons \cite{PhysRev.177.2426,Bell:1969ts}.
The anomalous trace is only dependent on the local curvature of spacetime and is not affected by the quantum state of the quantum fields. In a four-dimensional spacetime, it can be expressed as
\begin{equation}
	g^{\mu \nu} \langle T_{\mu \nu} \rangle = \frac{\beta}{2} \, C^2 - \frac{\alpha}{2}\mathcal{G},
	\label{eq:trace}
\end{equation}
where $C^2 = \frac{1}{3}R^2 - 2 R_{\mu \nu} R^{\mu \nu} + R_{\mu \nu \rho \sigma} R^{\mu \nu \rho \sigma}$ is the square of the Weyl tensor, $\mathcal{G} = R^2 - 4 R_{\mu \nu} R^{\mu \nu} + R_{\mu \nu \rho \sigma} R^{\mu \nu \rho \sigma}$ the Gauss-Bonnet scalar, and $\beta$ and $\alpha$ are coupling constants.
The Gauss-Bonnet and Weyl contributions to the anomaly are named type-A and type-B anomalies \cite{Deser:1993yx}, respectively.

Unlike most other modifications of General Relativity, incorporating the contributions from the trace anomaly into the low-energy effective field theory of gravity is essential, and is expected to result in macroscopic effects (see the discussion in Refs. \cite{Mottola:2022tcn,Mottola:2016mpl,Mottola:2006ew}).
To account for the effects of the trace anomaly on the spacetime geometry, a semiclassical approach can be used. This involves treating spacetime classically while considering the backreaction of quantum fields. The semiclassical Einstein equations take the form
\begin{equation}
	R_{\mu \nu} - \frac{1}{2} g_{\mu \nu} R = 8\pi G\, \langle T_{\mu \nu} \rangle,
	\label{eq:EinsteinEqs}
\end{equation}
where contributions from other matter sources were neglected.

Investigating the backreaction of quantum fields is often challenging, primarily because the expectation value of the renormalized stress-energy tensor $\langle T_{\mu\nu}\rangle$ is typically unknown. Exceptions exist, such as in the case of a homogeneous and isotropic spacetime, where the trace anomaly \eqref{eq:trace} completely determines the renormalized stress-energy tensor \cite{Hawking:2000bb,Starobinsky:1980te,Herzog:2013ed}. However, in a static and spherically symmetric system, the renormalized stress-energy tensor can only be determined up to an arbitrary function of position \cite{Christensen:1977jc,Ho:2018fwq,Abedi:2015yga,Beltran-Palau:2022nec}.
To overcome this difficulty, Ref. \cite{Cai:2009ua} imposed an additional condition: that the geometry should depend solely on one free function. In light of the field equations \eqref{eq:EinsteinEqs}, this assumption is equivalent to imposing an additional equation of state on the stress-energy tensor. By doing so and considering only the type-A anomaly ($\beta=0$), Ref. \cite{Cai:2009ua} was able to fully determine the renormalized stress-energy tensor. This approach led to the successful derivation of analytic static and spherically symmetric black hole solutions to the semiclassical Einstein equations \eqref{eq:EinsteinEqs}. 
Interestingly, the black hole solutions obtained in Ref. \cite{Cai:2009ua} exhibit a logarithmic correction to their entropy. This is consistent with the expectation that the leading-order quantum corrections to black hole entropy are logarithmic \cite{Page:2004xp,Sen:2012dw}. Similar quantum corrections were also studied, e.g., in Refs. \cite{Tomozawa:2011gp,Cognola:2013fva,Cai:2014jea,Calza:2022szy,Tsujikawa:2023egy} and in the scope of the recently formulated 4D-Einstein-Gauss-Bonnet theory \cite{Glavan:2019inb,Lu:2020iav,Fernandes:2020nbq,Hennigar:2020lsl,Fernandes:2022zrq}.

In this Letter, we will tackle the challenge of studying the backreaction of quantum fields sourcing the trace anomaly in a stationary and axially-symmetric setting. To achieve this, we will adopt a Kerr-Schild ansatz \cite{KerrSchild} for the metric, which will allow us to have just one free function in the geometry. By doing so, we can fully determine the renormalized stress-energy tensor, and derive exact analytic stationary and axially-symmetric black hole solutions to semiclassical Einstein equations \eqref{eq:EinsteinEqs}. These solutions present quantum-corrected alternatives to the Kerr black hole \cite{Kerr:1963ud}. Throughout this Letter we will use units where $8\pi G=c=1$.

%%%%%%%%%%%%%%%%%%%%%%%%%%%%%%%%%
\noindent{\bf{\em Setup.}}
%\section{Setup}
%%%%%%%%%%%%%%%%%%%%%%%%%%%%%%%%%
We are interested in stationary and axially-symmetric spacetimes. The simplest choice for this class of metrics follows from a Kerr-Schild ansatz \cite{KerrSchild,Stephani:2003tm}
\begin{equation}
	ds^2 = ds^2_{\mathrm{flat}} + H(\mathbf{x}) \left(l_\mu dx^\mu\right)^2,
	\label{eq:KerrSchild}
\end{equation}
where $ds^2_{\mathrm{flat}}$ is the line-element of Minkowski spacetime, $H(\mathbf{x})$ a scalar function, and $l^\mu$ is the tangent vector to a shear-free and geodesic null congruence. In ingoing Kerr-like coordinates $x^\mu=(v,r,\theta,\varphi)$, the line-element we will employ reads
\begin{equation}
	\begin{aligned}
		ds^2 =& -\left(1-\frac{2 r \mathcal{M}(r,\theta)}{\Sigma}\right) \left(dv - a \sin^2 \theta d\varphi\right)^2\\&+2\left(dv-a \sin^2 \theta d\varphi\right)\left(dr-a \sin^2 \theta d\varphi\right)\\&+\Sigma \left(d\theta^2 + \sin^2 \theta d\varphi^2\right),
	\end{aligned}
	\label{eq:metric}
\end{equation}
where we have taken $H(\mathbf{x}) = 2r \mathcal{M}(r,\theta)/\Sigma$ in Eq. \eqref{eq:KerrSchild}, $\Sigma = r^2+a^2 \cos^2\theta$, and where $\mathcal{M}(r,\theta)$ is the mass function defining the spacetime.
When we set $\mathcal{M}(r,\theta)=M$, where $M$ is a constant that represents the ADM mass, we recover the Kerr metric \cite{Kerr:1963ud}, which is the stationary and axially-symmetric black hole solution to the vacuum Einstein equations.

The expectation value of the renormalized stress-energy tensor should be compatible with the metric \eqref{eq:metric} and its symmetries, and it must reduce to the anisotropic stress-energy tensor employed in Ref. \cite{Cai:2009ua} in the static limit. Moreover, it is important to note that the trace anomaly in Eq. \eqref{eq:trace} is insensitive to traceless contributions to the stress-energy tensor.
The most general non-trivial ansatz that satisfies these constraints is \cite{Ibohal:2004kk}
\begin{equation}
	\begin{aligned}
		\langle T_{\mu \nu} \rangle =& 2 \left(\rho + p_t\right) \, l_{(\mu} n_{\nu)} + p_t \, g_{\mu \nu} \\& + \mu  \, l_\mu l_\nu - 4 \operatorname{Re}\left\{ \omega \, l_{(\mu} \overline{m}_{\nu)} \right\},
	\end{aligned}
	\label{eq:stress-energy}
\end{equation}
where $l^\mu$, $n^\mu$ and $m^\mu$ form a complex null tetrad $e^{\phantom{\mu}\mu}_{\hat \mu} = \left(l^\mu, n^\mu, m^\mu, \overline{m}^\mu\right)$ (given in the \textit{Supplemental Material}) and obey the usual relations $l_\mu n^\mu = -m_\mu \overline{m}^\mu = -1$, with all other inner products vanishing.
The quantities $\rho$, $p_t$, $\mu$, and $\omega$ are all functions of the coordinates $r$ and $\theta$, and the brackets denote symmetrization in the usual way. The energy density and transverse pressure of the quantum fields are given by $\rho$ and $p_t$, respectively, and both are Lorentz invariant. On the other hand, $\mu$ and $\omega$ are not Lorentz invariant and are associated with a traceless contribution. Stress-energy tensors of a similar form to Eq. \eqref{eq:stress-energy} have been previously used to construct rotating radiating black holes \cite{Ibohal:2004kk,Carmeli:1975kg,PhysRevD.1.3220}, and their matter content interpreted as a combination of an anisotropic fluid and a null string fluid \cite{Gurses:1975vu,10.1093/ptep/ptab116}.

The trace of the semiclassical Einstein equations \eqref{eq:EinsteinEqs} together with the trace anomaly \eqref{eq:trace} yields a useful relation that depends solely on the geometry
\begin{equation}
	R + \frac{\beta}{2} C^2 - \frac{\alpha}{2}\mathcal{G} = 0.
	\label{eq:trace_eq}
\end{equation}

%%%%%%%%%%%%%%%%%%%%%%%%%%%%%%%%%
\noindent{\bf{\em Solving the semiclassical Einstein equations.}}
%\section{Solving the semiclassical Einstein equations}
%%%%%%%%%%%%%%%%%%%%%%%%%%%%%%%%%
To obtain exact analytical solutions, we will set $\beta = 0$ for now, focusing on the type-A anomaly. The non-vanishing $\beta$ case will be addressed later. 
We have five unknowns, namely $\rho$, $p_t$, $\mu$, $\omega$ and $\mathcal{M}$, and four independent equations provided by the semiclassical Einstein equations \eqref{eq:EinsteinEqs} (shown in the \textit{Supplemental Material} for the metric ansatz \eqref{eq:metric}) and the covariant conservation of the stress-energy tensor $\nabla_\mu \langle T^{\mu \nu}\rangle = 0$. One additional equation is needed to close the system, which can be chosen as either the trace anomaly \eqref{eq:trace} or the trace equation \eqref{eq:trace_eq}. To expedite the solution-finding process, we will take a shortcut by solving the trace equation \eqref{eq:trace_eq} and then determining the stress-energy tensor quantities from the semiclassical Einstein equations. Note that we can only take this shortcut because our ``equation of state'', the trace anomaly, is such that the trace of the renormalized stress-energy tensor depends only on the metric. An alternative approach, which is equally valid, involves fully determining the stress-energy tensor using its covariant conservation and the trace anomaly \eqref{eq:trace}, and subsequently employing it in the semiclassical Einstein equations to obtain a solution as done e.g. in Refs. \cite{Hawking:2000bb,Cai:2009ua}. Both approaches lead to the same outcome.

For our spacetime ansatz \eqref{eq:metric}, it can be shown that the following relations hold
\begin{equation}
	\Sigma R = 2 \partial_r^2 \left(r \mathcal{M} \right), \quad \Sigma \mathcal{G} = 8 \partial_r^2 \left(\frac{r^2 \mathcal{M}^2 \xi}{\Sigma^3} \right),
\end{equation}
where $\xi = r^2 - 3a^2 \cos^2\theta$.
To solve the trace equation \eqref{eq:trace_eq}, we use these expressions for the Ricci and Gauss-Bonnet scalars, obtaining the following equivalent condition
\begin{equation}
	\partial_r^2 \left( r \mathcal{M} - 2\alpha \frac{r^2 \mathcal{M}^2 \xi}{\Sigma^3} \right) = 0.
	\label{eq:scalars}
\end{equation}
This equation can trivially be solved by integrating twice and solving algebraically for $\mathcal{M}$, resulting in the general solution
\begin{equation}
	\mathcal{M}(r,\theta) = \frac{2 \left(M - \frac{q}{2r} \right)}{1 \pm \sqrt{1-\frac{8\alpha r \xi}{\Sigma^3} \left(M-\frac{q}{2r}\right) }},
	\label{eq:Msol}
\end{equation}
where $M \equiv M(\theta)$ and $q \equiv q(\theta)$ are integration constants. Using the Einstein equations \eqref{eq:EinsteinEqs} we fully determine the renormalized stress-energy tensor by obtaining the profiles for $\rho$, $p_t$, $\mu$ and $\omega$.

%%%%%%%%%%%%%%%%%%%%%%%%%%%%%%%
\noindent{\bf{\em Nature of the stress-energy tensor.}}
%\section{Nature of the stress-energy tensor}
%%%%%%%%%%%%%%%%%%%%%%%%%%%%%%%
To better understand the physical origins of the stress-energy tensor \eqref{eq:stress-energy} that sources the semiclassical Einstein equations with the type-A trace anomaly \eqref{eq:trace}, one possible approach is to consider an appropriate effective action that captures the trace anomaly in a gravitational theory \cite{Riegert:1984kt,Komargodski:2011vj,Mazur:2001aa,Mottola:2022tcn,Gabadadze:2023quw,Tsujikawa:2023egy,Fernandes:2021dsb}.
Such an effective action can be written as $S = \int d^4x \mathcal{L}$, where the Lagrangian density $\mathcal{L}$ takes the form \cite{Gabadadze:2023quw,Fernandes:2021dsb}
\begin{equation}
	\begin{aligned}
		&\frac{\mathcal{L}}{\sqrt{-g}} = \frac{R}{2} - \frac{1}{2}\left(\partial \phi\right)^2 - \frac{\phi^2}{12} R +\frac{\alpha}{2}\bigg[\ln(\phi) \mathcal{G}\\&
		- \frac{4 G^{\mu \nu} \partial_\mu \phi \partial_\nu \phi}{\phi^2} - \frac{4\Box \phi \left(\partial \phi\right)^2}{\phi^3} + \frac{2 \left(\partial \phi\right)^4}{\phi^4} \bigg],
	\end{aligned}
	\label{eq:effaction}
\end{equation}
modulo other conformally invariant terms. Observe that the Einstein-Hilbert term is supplemented with a conformally coupled scalar field \cite{Fernandes:2021dsb,Ayon-Beato:2023bzp,Babichev:2023dhs,Babichev:2023rhn}. It can be shown that the on-shell trace of the stress-energy tensor of the scalar gives the type-A trace anomaly \eqref{eq:trace} \cite{Fernandes:2021dsb}. The theory described by the above action belongs to the Horndeski class \cite{Horndeski:1974wa,Fernandes:2021dsb} and is intimately connected to the well-defined scalar-tensor formulations of the 4D-Einstein-Gauss-Bonnet class of theories \cite{Glavan:2019inb,Lu:2020iav,Fernandes:2020nbq,Hennigar:2020lsl,Fernandes:2022zrq}.

The field equations resulting from the action \eqref{eq:effaction} are highly complex and challenging to solve in a stationary and axially-symmetric setting. They are presented in the \textit{Supplemental Material}.
However, we have been able to obtain a simple solution to these field equations by considering the metric ansatz \eqref{eq:metric}. As discussed earlier, the mass function in Eq. \eqref{eq:Msol} solves the trace condition of the theory, $R-\frac{\alpha}{2}\mathcal{G}=0$, and remarkably the remaining field equations can be solved with a constant but non-zero scalar field $\phi = \pm \sqrt{6}$\footnote{It is anticipated that a strong coupling issue may arise for the scalar field at this particular constant value \cite{Babichev:2023dhs}, leading to a divergent effective Newton's constant. Nevertheless, it is plausible that alternative scalar profiles exist, resulting in the same geometry, similar to what has been observed in the static and spherically symmetric case \cite{Fernandes:2021dsb} and the slowly-rotating case \cite{Charmousis:2021npl, Gammon:2022bfu}. The exploration of such possibilities is deferred to future research. Nonetheless, this toy model serves as an illustrative example that can help clarify the nature of the matter/energy content of the stress-energy tensor, given its unconventional form.}.
As a result, one of the possible interpretations for the renormalized stress-energy tensor \eqref{eq:stress-energy} is that it corresponds to that of a scalar field sourcing the anomaly. Hairy black hole solutions supported by a constant conformally coupled scalar field have been previously studied, e.g., in Ref. \cite{Astorino:2013sfa}.

The fulfillment of energy conditions \cite{Martin-Moruno:2017exc} is crucial for the viability of classical solutions to the Einstein equations. However, there is a growing body of evidence, derived from both experimental and theoretical sources, suggesting that quantum effects are likely to lead to the widespread violation of some or potentially all of these conditions (see Refs. \cite{Casimir:1948dh,PhysRevD.36.1065,Visser:1997tq,Burinskii:2001bq,Visser:1997gf,Santiago:2021aup,Martin-Moruno:2017exc} and references therein, as well as Refs. \cite{Visser:1994jb,PhysRevD.44.403} for explicit examples of energy condition violations due to the trace anomaly). Due to the purely quantum nature of $\langle T_{\mu \nu} \rangle$, it is anticipated that our system will exhibit violations of at least some energy conditions. In fact, a straightforward analysis demonstrates that violations, e.g., of the weak energy condition, are already generically present in static solutions\footnote{In the static case, in an orthonormal frame, the stress-energy tensor assumes a simple diagonal form $\langle T_{\hat{\mu} \hat{\nu}} \rangle = \mbox{diag}\left(\rho,-\rho,p_t,p_t\right)$.}. A detailed analysis of these violations in the rotating case can be done using the results of Ref. \cite{Ibohal:2004kk}, but further exploration of these considerations is reserved for future investigations.

%%%%%%%%%%%%%%%%%%%%%%%%%%%%%%%
\noindent{\bf{\em Properties of the solution.}}
%\section{Properties of the solution}
%%%%%%%%%%%%%%%%%%%%%%%%%%%%%%%
The mass function in Eq. \eqref{eq:Msol} defines the spacetime, and it contains two integration constants, $M(\theta)$ and $q(\theta)$, and two different branches of solutions (as indicated by the $\pm$ sign in the denominator).
Imposing the boundary condition that the mass function should approach a constant at infinity, which is interpreted as the ADM mass, leads us to choose the solution with the plus sign as the physical one, and constant $M$.

The integration constant $q(\theta)$ can be understood through the falloff behavior of the energy density and transversal pressure near spatial infinity, which is given by $\rho \approx p_t \approx q(\theta)/r^4 + \mathcal{O}(r^{-5})$. This falloff is characteristic of a conformal field theory with $U(1)$ squared conserved charge proportional to $q$. Therefore, $q$ is interpreted as an artifact of considering all possible suitable traceless contributions to the stress-energy tensor and plays a role similar to that of an electric charge \cite{Cai:2009ua}. For simplicity, we set $q=0$ from now on. The same conditions on the integration constants could be obtained by demanding that the Kerr metric is recovered as $\alpha$ approaches zero.

From the asymptotic behavior of the metric component $g_{v\phi}$, we can determine the angular momentum (per unit mass), which is given by $a$. This solution is algebraically special, namely Petrov type II in the Petrov classification scheme \cite{Petrov:2000bs,Stephani:2003tm}. Moreover, this solution does not satisfy the circularity conditions \cite{Delaporte:2022acp,Xie:2021bur} in general\footnote{This is true for non-vanishing $\alpha$ and $a$. If either is zero, the solution is Petrov type D and obeys the circularity conditions.} because of the angular dependency of the function $\mathcal{M}$ \cite{Ayon-Beato:2015nvz}\footnote{Other examples of non-circular black hole solutions can be found e.g. in Refs. \cite{Anson:2020trg,BenAchour:2020fgy,Minamitsuji:2020jvf}.}.

Although the solution of Ref. \cite{Cai:2009ua} can be obtained in the limit where $a$ approaches zero, it is not possible to derive the solution \eqref{eq:Msol} directly by applying the Newman-Janis algorithm \cite{Newman:1965tw} to the static solution in Ref. \cite{Cai:2009ua}. This algorithm is known to fail in modified theories of gravity \cite{Hansen:2013owa,Kamenshchik:2023bzc}, and in our case, it results in a geometry that violates the trace condition \eqref{eq:trace_eq}. Our findings demonstrate the importance of not blindly applying the Newman-Janis algorithm. Additionally, in the limit of slow rotation, we also recover the slowly-rotating black holes of 4D-Einstein-Gauss-Bonnet gravity \cite{Charmousis:2021npl,Gammon:2022bfu}.

The physical singularities of the spacetime \eqref{eq:Msol} can be analyzed by computing, for example, the Ricci scalar in Eq. \eqref{eq:scalars}. We can observe the presence of two singularities. The first one is the usual ring singularity, located at $\Sigma = 0$, i.e., $r=0$ and $\theta = \pi/2$. The second singularity is introduced by Gauss-Bonnet quantum effects and is located where the quantity inside the square-root in the mass function \eqref{eq:Msol} vanishes
\begin{equation}
	1-\frac{8\alpha r \xi}{\Sigma^3} M = 0,
	\label{eq:singularity}
\end{equation}
where $r$ is to be evaluated at $r_s(\theta)$, the location of the singularity. This condition has to be solved numerically, except for $\theta=\pi/2$, where we obtain $r_s(\pi/2) = 2(M \alpha)^{1/3}$. It is worth noting that finite radius singularities are a common feature of theories containing Gauss-Bonnet terms, and have been extensively studied, for instance, in Ref. \cite{Fernandes:2022kvg}.

\begin{figure}
	\centering
	\includegraphics[width=\columnwidth]{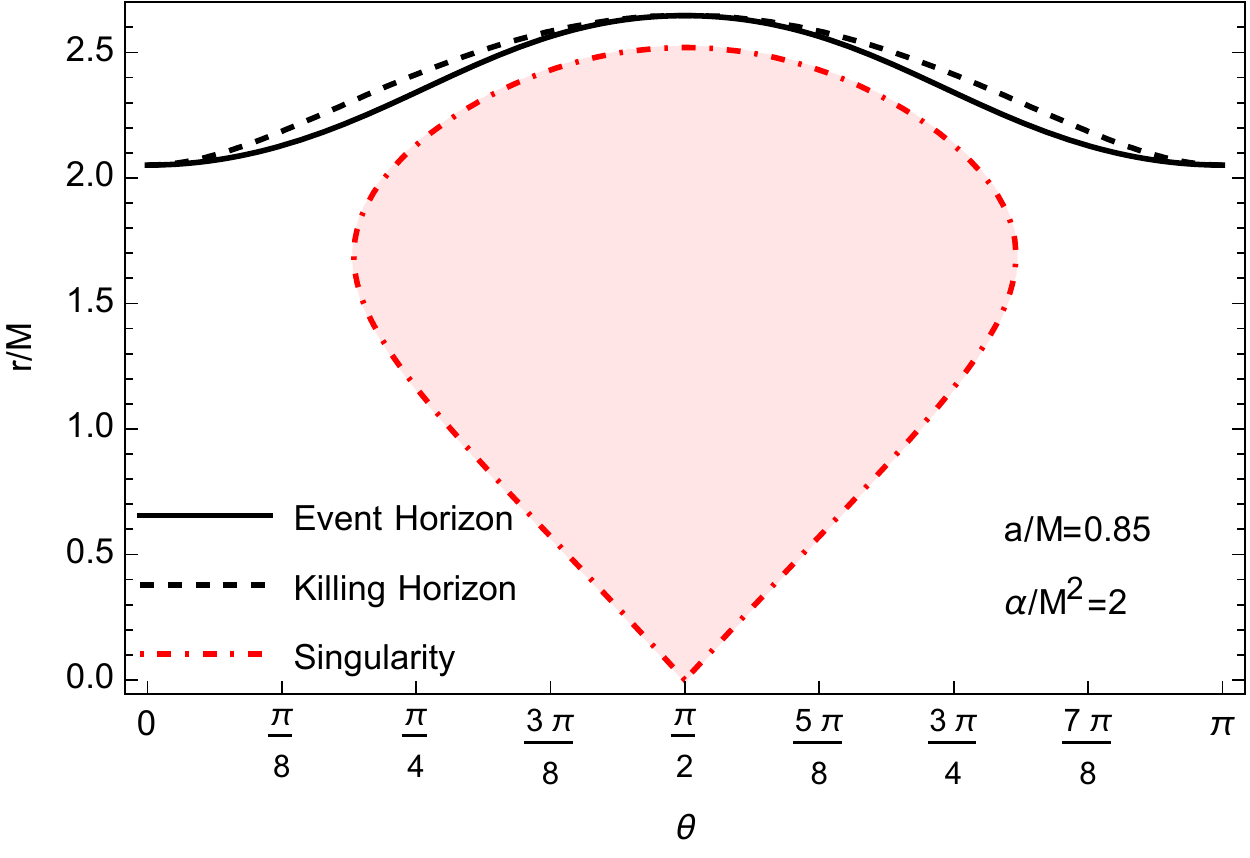}
	\caption{Profile for the event (black solid line) and Killing (black dashed line) horizons coordinate location as a function of the angular coordinate $\theta$ for a black hole with $a/M=0.85$ and $\alpha/M^2=2$. The red dot-dashed line indicates the location of the curvature singularity in Eq. \eqref{eq:singularity}. The singularity is hidden by the event horizon.}
	\label{fig:horizons}
\end{figure}

To confirm that the solution \eqref{eq:Msol} describes a black hole spacetime, we need to investigate the existence of event and Killing horizons. In vacuum General Relativity, the rigidity theorem holds \cite{Hawking:1973uf}, and both types of horizons coincide. However, this is not necessarily the case for the solution \eqref{eq:Msol}. The coordinate location of the event horizon $r_H$ depends on the coordinate $\theta$ and is given by the solution to the differential equation \cite{Johannsen:2013rqa,Eichhorn:2021iwq,Anson:2020trg}
\begin{equation}
	\left[\partial_\theta r_H(\theta)\right]^2 + \Delta|_{r=r_H(\theta)} = 0,
	\label{eq:EH}
\end{equation}
where
\begin{equation}
	\Delta = r^2 + a^2 - 2r \mathcal{M}(r,\theta).
	\label{eq:Delta}
\end{equation}
The location of the Killing horizon is given by the solution to the equation $\Delta = 0$ \cite{Johannsen:2013rqa,Eichhorn:2021iwq}.

To solve the differential equation \eqref{eq:EH}, we used a pseudospectral method by expanding $r_H(\theta)$ in a spectral series of even cosines, taking into account the symmetries of the problem, $\partial_\theta r_H\left(0\right) = \partial_\theta r_H\left(\pi/2\right) = 0$. We verified the existence of regular event horizons for the solution \eqref{eq:Msol} in a domain, confirming that it is a black hole spacetime. The coordinate location of the event and Killing horizons both depend on the angular coordinate $\theta$, thereby deviating from spherical symmetry, and coincide at the poles and equator. We remark that this angular dependence is a consequence of non-circularity.
As an example, in Fig. \ref{fig:horizons}, we plot the location of the event horizon, Killing horizon, and that of the curvature singularity in Eq. \eqref{eq:singularity} for a solution with $a/M=0.85$ and $\alpha/M^2=2$, where we observe that the singularity is hidden inside the event horizon.

\begin{figure}[]
	\centering
	\includegraphics[width=\columnwidth]{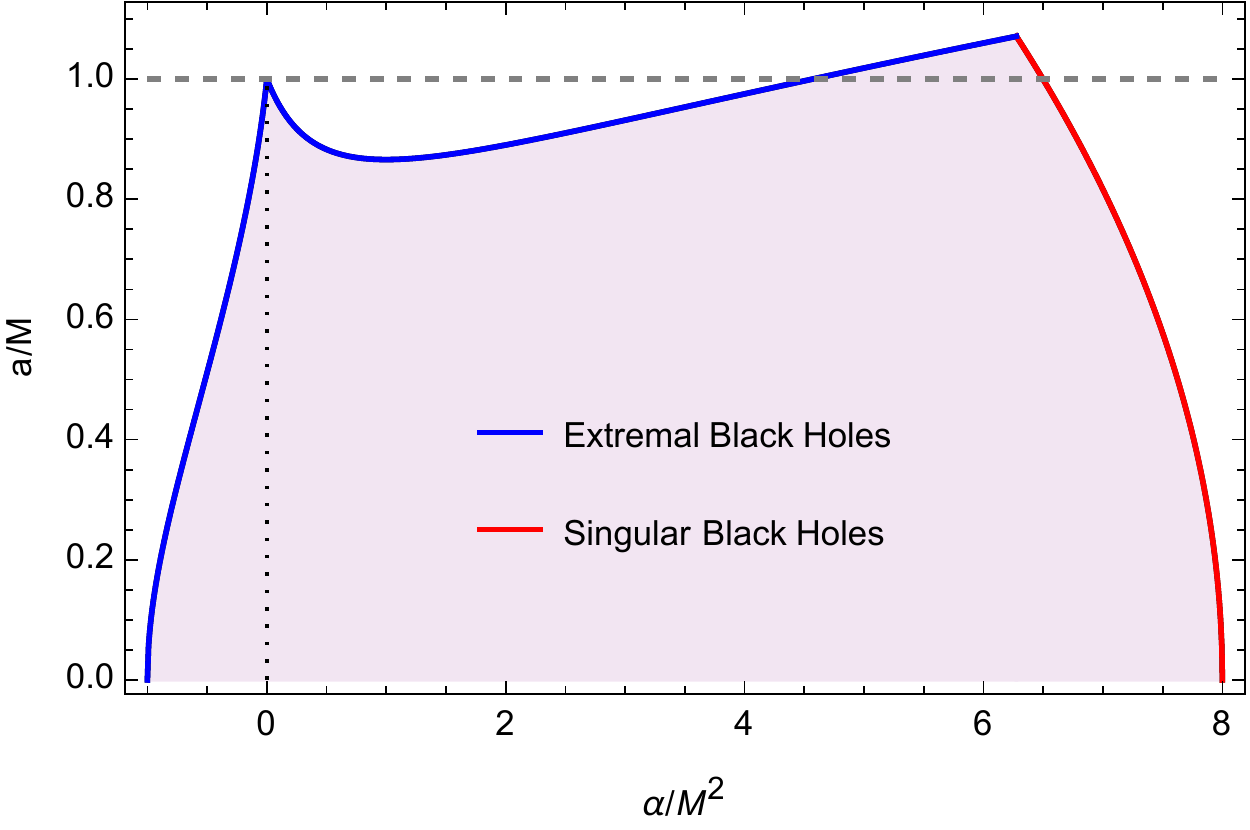}
	\caption{Domain of existence for black hole solutions in the $(\alpha/M^2, a/M)$ plane. The shaded region represents the parameter space in which black hole solutions exist. The boundary of this region is formed by extremal (blue) and singular black holes (red).}
	\label{fig:domain}
\end{figure}

In Fig. \ref{fig:domain}, the shaded region shows the domain of existence of black holes.
For values $-1 \leq \alpha/M^2 \lesssim 6.2754$, the domain is bounded by extremal black holes.
As we approach the blue line for positive (negative) couplings, the event and inner horizons, the two roots of $\Delta$, overlap at the poles (equator). However, for higher values of $\alpha/M^2$, we observe the overlap of the location of the event horizon and the curvature singularity \eqref{eq:singularity} at the equator as we approach the red line, resulting in a singular solution. We found violations of the Kerr bound, $a/M\leq 1$, for values $4.5698 \lesssim \alpha/M^2 \lesssim 6.4976$, with a maximum value of $a/M \approx 1.07109$ for $\alpha/M^2 \approx 6.2754$. This is the point at which the extremal and singular black hole branches come together. This kind of behavior was also observed in previous studies of black holes in Einstein-dilaton-Gauss-Bonnet gravity \cite{Kleihaus:2015aje}.

The region where the normalized timelike Killing vector field at infinity $\partial_v$ becomes null is known as the ergoregion. In the spacetime described by equation \eqref{eq:metric}, ergoregions are located at points where the condition $g_{vv} = 0$ is satisfied. We have confirmed that ergoregions are present in the entire domain of existence of black holes with the usual $S^2$ topology.

%%%%%%%%%%%%%%%%%%%%%%%%%%%%%%%
\noindent{\bf{\em Addressing the full anomaly.}}
%\section{Addressing the full anomaly}
%%%%%%%%%%%%%%%%%%%%%%%%%%%%%%%
When $\beta$ is non-vanishing, the Weyl tensor introduces non-linearities into the system and closed-form solutions are not known, even in the static and spherically symmetric case \cite{Cai:2009ua,Cognola:2013fva}. To address the full anomaly, we take $\beta = k \alpha$ for some proportionality constant $k \sim \mathcal{O}(1)$ and solve the trace condition \eqref{eq:trace_eq} perturbatively in powers of $\alpha/M^2$. There are two special cases to consider: when $k=0$, we recover the results and solution presented in previous sections; when $k=1$, all terms proportional to the square of the Riemann tensor cancel in Eq. \eqref{eq:trace_eq}, leaving us with a trace condition depending solely on the Ricci tensor and scalar. In this case, it follows that the (Ricci-flat) Kerr solution $\mathcal{M}(r,\theta) = M$ solves the field equations for vanishing stress-energy tensor. For general $k$, we can consider perturbative solutions in an expansion in $\alpha/M^2$ away from the Kerr solution\footnote{An expansion in powers of $(1-k)$ away from the Kerr metric could also be considered.}
\begin{equation}
	\mathcal{M}(r,\theta) = M + \sum_{n=1}^\infty \left(\frac{\alpha}{M^2}\right)^n \tilde{\mathcal{M}}_n(r,\theta).
\end{equation}
By solving the trace condition \eqref{eq:trace_eq} order-by-order, we obtain the following perturbative solution to order $\left(\alpha/M^2\right)^2$
\begin{equation}
	\tilde{\mathcal{M}}_1=\frac{2 (1-k) M^4 r \xi}{\Sigma^3}, \quad \tilde{\mathcal{M}}_2 = \frac{8 (1-k)^2 M ^7 r^2 \xi^2}{\Sigma^6}.
	\label{eq:perturb}
\end{equation}
The perturbative solutions retain many of the characteristics of the solution \eqref{eq:Msol}, such as a function $\mathcal{M}$ that is dependent on both $r$ and $\theta$, leading to non-circularity and an event horizon that is not spherically symmetric.

It is interesting to note that in the static limit of the perturbative solution \eqref{eq:perturb}, a direct application of the first law of thermodynamics \cite{Cai:2009ua} shows that to leading order in the coupling, the black hole entropy $\mathcal{S}$ obeys
\begin{equation}
	\mathcal{S} = \frac{A_H}{4} - 2\pi \alpha \left(1-k\right) \ln \left(\frac{A_H}{A_0}\right),
\end{equation}
where $A_H$ is the event horizon area, and $A_0$ is a (squared) length scale to be set by the ultraviolet complete theory.
This means that the leading-order logarithmic corrections to the black hole entropy are a general prediction of trace anomaly quantum-corrections and are not limited to the $\beta=0$ case \cite{Cai:2009ua}.

%%%%%%%%%%%%%%%%%%%%%%%%%%%%%%%
\noindent{\bf{\em Discussion.}}
%\section{Discussion}
%%%%%%%%%%%%%%%%%%%%%%%%%%%%%%%
The stationary and axially-symmetric solution \eqref{eq:Msol} found in this Letter describes a spinning black hole that incorporates trace anomaly quantum corrections. This solution possesses distinctive characteristics, including a non-spherically-symmetric event horizon and violations of the Kerr bound.
It provides an alternative to the traditional Kerr black hole geometry and serves as an analytical black hole solution within the framework of scalar-Gauss-Bonnet gravity \cite{Doneva:2017bvd,Silva:2017uqg,Antoniou:2017acq,Herdeiro:2020wei,Berti:2020kgk,Cunha:2019dwb,Kleihaus:2015aje}, broadly defined. These unique features make it a subject worthy of further investigation and study. Suggested follow-up work includes studying the quasinormal modes, light rings, innermost stable circular orbit, and black hole mechanics/thermodynamics in-depth. Regarding the latter, it is straightforward to verify that in the static limit, the surface gravity remains constant on the event horizon and that the first law of black hole mechanics, $dM = T d\mathcal{S}$, holds. However, a more meticulous analysis is necessary for the rotating case due to the angular dependence of the horizons on the coordinate $\theta$ resulting from non-circularity, and the fact that the Killing and event horizons do not coincide (except at the poles and equator).

Additionally, it would be important to study the image features and shadow of this black hole geometry in light of observations by the Event Horizon Telescope collaboration \cite{EventHorizonTelescope:2019dse,EventHorizonTelescope:2022wkp}. The non-circularity of the spacetime is anticipated to give rise to interesting image features \cite{Delaporte:2022acp,Eichhorn:2021iwq,Eichhorn:2021etc}. Another direction for new analytic solutions would be to generalize this solution to include the effects of a cosmological constant \cite{Cai:2014jea}.

%%%%%%%%%%%%%%%%%%%%%%%%%%%%%%%
\noindent{\bf{\em Acknowledgments.}}
%\section*{Acknowledgments.}
%%%%%%%%%%%%%%%%%%%%%%%%%%%%%%%
P.F. is supported by a Research Leadership Award from the Leverhulme Trust. P.F. thanks Tiago França for valuable discussions that led to this project, as well as Clare Burrage and Astrid Eichhorn for their insightful feedback and comments on a previous version of the manuscript.

\bibliography{biblio}

%\clearpage
\onecolumngrid
\newpage
\appendix
\renewcommand{\appendixname}{Supplemental Material}
\section{Complex null tetrad and the Einstein equations}
We define the complex null tetrad $e^{\phantom{\mu}\mu}_{\hat \mu} = \left(l^\mu, n^\mu, m^\mu, \overline{m}^\mu\right)$, where
\begin{equation}
	l^\mu = \{0,1,0,0\}, \qquad n^\mu = -\frac{1}{\Sigma} \left\{ (r^2+a^2), \Delta/2, 0 , a \right\}, \qquad m^\mu = \frac{1}{\sqrt{2} \sigma} \left\{ i a \sin \theta, 0, -1, i/\sin \theta \right\},
	\label{eq:nullbasis}
\end{equation}
with the overline denoting complex conjugation, $\sigma = r + i a \cos \theta$, and $\Delta$ is defined in Eq. \eqref{eq:Delta}.
In the null basis \eqref{eq:nullbasis} the non-vanishing components of the stress-energy tensor take the simple form
\begin{equation}
	\langle T_{\hat{v} \hat{r}} \rangle = \rho, \qquad \langle T_{\hat{r} \hat{r}} \rangle = \mu, \qquad \langle T_{\hat{r} \hat{\theta}} \rangle = \overline{\langle T_{\hat{r} \hat{\varphi}}} \rangle = \omega, \qquad \langle T_{\hat{\theta} \hat{\varphi}} \rangle = p_t,
\end{equation}
and the non-trivial components of the Einstein tensor are
\begin{equation}
	\begin{aligned}
		&G_{\hat{v} \hat{r}} = \frac{2r^2 \partial_r \mathcal{M}}{\Sigma^2}, \qquad
		G_{\hat{r} \hat{r}} = -\frac{r \left(\cot \theta \partial_\theta + \partial_\theta^2\right)\mathcal{M}}{\Sigma^2},\\&
		G_{\hat{r} \hat{\theta}} = \overline{G_{\hat{r} \hat{\varphi}}} = \frac{\left(-\sigma \partial_\theta + r \overline{\sigma} \partial_r \partial_\theta  \right)\mathcal{M}}{\sqrt{2} \Sigma^2}, \qquad
		G_{\hat{\theta} \hat{\varphi}} = -\frac{\left(2 a^2 \cos ^2\theta  \partial_r+r \Sigma  \partial_r^2 \right)\mathcal{M}}{\Sigma^2}.
	\end{aligned}
\end{equation}

\section{Field equations for the conformally coupled scalar field theory}
The conformally coupled scalar field theory in Eq. \eqref{eq:effaction} has field equations given by \eqref{eq:EinsteinEqs} with \cite{Fernandes:2021dsb}
%\begin{widetext}
\begin{equation}
	\begin{aligned}
		\langle T_{\mu \nu} \rangle = & \alpha \bigg[ 2G_{\mu \nu} \left(\partial \varphi\right)^2-4 {}^*R^*_{\mu \alpha \nu \beta}\left(\nabla^\alpha \varphi \nabla^\beta \varphi - \nabla^\beta \nabla^\alpha \varphi \right) + 4\left(\nabla_\alpha \varphi \nabla_\mu \varphi - \nabla_\alpha \nabla_\mu \varphi\right) \left(\nabla^\alpha \varphi \nabla_\nu \varphi - \nabla^\alpha \nabla_\nu \varphi\right)\\
		&+4\left(\nabla_\mu \varphi \nabla_\nu \varphi - \nabla_\nu \nabla_\mu \varphi\right) \Box \varphi +g_{\mu \nu} \Big(2\left(\Box \varphi \right)^2 - \left( \partial \varphi\right)^4 + 2\nabla_\beta \nabla_\alpha \varphi \left(2\nabla^\alpha \varphi \nabla^\beta \varphi - \nabla^\beta \nabla^\alpha \varphi \right) \Big) \bigg] \\&
		+ \frac{1}{6} e^{2\varphi} \left[G_{\mu \nu} + 2\nabla_\mu \varphi \nabla_\nu \varphi - 2\nabla_\mu \nabla_\nu \varphi +g_{\mu \nu} \left( 2\Box \varphi + \left(\partial \varphi\right)^2 \right)\right],
	\end{aligned}
\end{equation}
where, for simplicity, we used the auxiliary field $\varphi = \ln \phi$, and ${}^*R^*_{\mu \alpha \nu \beta}$ is the double-dual of the Riemann tensor. The scalar field equation resulting from the action \eqref{eq:effaction} is given by
\begin{equation}
	\begin{aligned}
		&-\frac{\alpha}{2} \bigg[ \mathcal{G} - 8 R^{\mu \nu}\nabla_{\mu} \varphi \nabla_{\nu} \varphi + 8G^{\mu \nu}\nabla_{\mu} \nabla_{\nu} \varphi+8\Box \varphi \left(\partial \varphi\right)^2 - 8\left(\nabla_{\mu} \nabla_{\nu} \varphi\right)^2 + 8\left(\Box \varphi\right)^2 + 16\left(\nabla_{\mu} \varphi \nabla_{\nu} \varphi\right)\left(\nabla^{\mu} \nabla^{\nu} \varphi\right) \bigg]\\&
		+\frac{1}{6}e^{2\varphi} \left[ R - 6 \Box \varphi -6 \left(\partial \varphi\right)^2 \right] = 0.
	\end{aligned}
\end{equation}
%\end{widetext}
Using the equation of motion for the scalar field, we can confirm that $g^{\mu \nu} \langle T_{\mu \nu} \rangle = -\frac{\alpha}{2}\mathcal{G}$. We can verify that $\phi = \pm \sqrt{6}$ solves the Einstein equations \eqref{eq:EinsteinEqs}. The equation of motion for the scalar field then simplifies to $R-\frac{\alpha}{2}\mathcal{G} = 0$, which can be solved by using the metric \eqref{eq:metric} and the mass function presented in Eq. \eqref{eq:Msol}.

\end{document}